# A Multichain based marketplace Architecture

**Muhammad Shoaib Farooq, Hamza Jamil, Hafiz Sohail Riaz,**
[1]Department of Computer Science, University of Management and Technology, Lahore, 54000, Pakistan

Corresponding author: Muhammad Shoaib Farooq (e-mail: shoaib.farooq@umt.edu.pk).

**ABSTRACT** A multichain non-fungible tokens (NFTs) marketplace is a decentralized platform where users can buy, sell, and trade NFTs across multiple blockchain networks by using cross communication bridge. In past most of NFT marketplace was based on singlechain in which NFTs have been bought, sold, and traded on a same blockchain network without the need for any external platform. The singlechain based marketplace have faced number of issues such as performance, scalability, flexibility and limited transaction throughput consequently long confirmation times and high transaction fees during high network usage. Firstly, this paper provides the comprehensive overview about NFT Multichain architecture and explore the challenges and opportunities of designing and implementation phase of multichain NFT marketplace to overcome the issue of single chain-based architecture. NFT multichain marketplace architecture includes different blockchain networks that communicate with each other. Secondly, this paper discusses the concept of mainchain interacting with sidechains which refers to multi blockchain architecture where multiple blockchain networks are connected to each other in a hierarchical structure and identifies key challenges related to interoperability, security, scalability, and user adoption. Finally, we proposed a novel architecture for a multichain NFT marketplace, which leverages the benefits of multiple blockchain networks and marketplaces to overcome these key challenges. Moreover, proposed architecture is evaluated through a case study, demonstrating its ability to support efficient and secure transactions across multiple blockchain networks and highlighting the future trends NFTs and marketplaces and comprehensive discussion about the technology.

## I. INTRODUCTION

In today's world where everything has become digitalized, technological progress means that discovery of new things and improved human facilities. Copyright and privacy are one of the major problems of the internet world. When we upload different types of content like images, mp3 or other documents the problem is that how we can define ownership. NFTs (Non-fungible tokens) emerged as a solution that addresses some of the problems that exist in today's internet world (Valeonti, et al., 2021). NFTs provide instant authentication and provenance. So, we can use NFTs to record and verify the ownership of unique digitals assets and we can claim over copyright and privacy. NFTs can be utilized to represent real estate, art works, pictures and music etc. There different characteristics of NFT shows in figure 1 that make its unique.

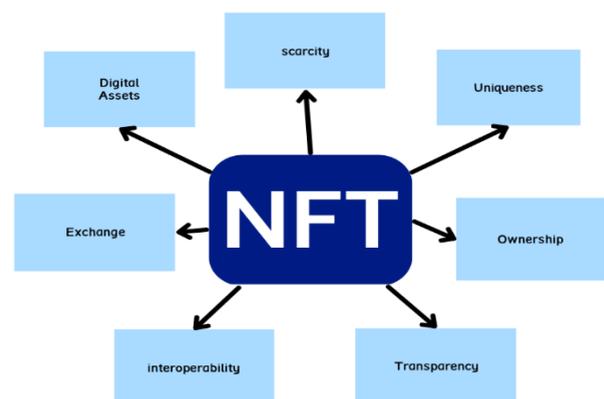

**Figure 1**. *NFT characteristics*

A multichain marketplace for NFTs is a platform where people can buy and sell NFTs. These marketplaces typically operate on multiple blockchain networks like Ethereum, binance smart chain and polygon and allow users to securely buy and sell NFTs using cryptocurrency (Kiong & Liew Voon, 2021). A multichain marketplace provides a way for NFTs collectors and enthusiasts to access a wide range of NFTs across multiple blockchain networks, without having to create accounts on each individual platform. This can increase liquidity in the NFTs market and make it easier for buyers and sellers to connect. Multichain NFTs marketplace network can be achieved by using bridges or cross chain protocols that enable interoperability between different blockchains (Huynh-The, et al., 2023). Bridge and cross communication mechanisms allows the transfer and communication of NFTs between different blockchain networks. These technologies provide enhanced flexibility in the NFTs space, enabling NFTs to move between different blockchain networks and facilitating the interaction between NFTs and smart contracts. Some characteristics of multichain networks shows in figure 2 for more understanding.

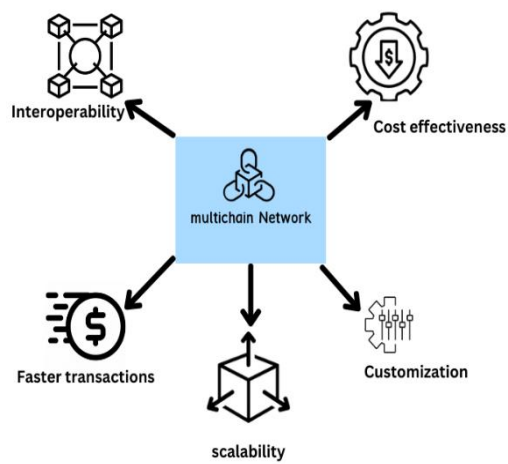

**Figure 2**.Multichain networks characteristics

Majority of NFT marketplaces are limited because of a singlechain based blockchain, which restricts the interoperability of NFTs and limits their stability performance and potential. However, this type of NFT marketplace face many challenges related to stability, performance and flexibility. (Zheng, X & Zhu, Yong & Si, & Xueming, 2019). If NFTs is created on Ethereum it cannot be traded on other blockchain networks such as Binance Smart Chain or Polygon. Which restricts the flexibility of NFTs and limits their potential value. And it also limits user attraction which means that these NFT marketplaces may have a limited number of users, particularly if the blockchain network is relatively new or has a smaller community. This thing can limit the liquidity and variety of NFTs available on the platform. However, to address these issues with singlechain NFT marketplace may have failed to explore partnerships with other blockchain networks to increase interoperability and invest in scalability solutions like layer 2 solutions.

To overcome this type of challenges, develop a multichain based NFT marketplace where users can buy, sell and trade NFTs across multiple blockchain networks. Unlike traditional marketplaces which may be limited to a single blockchain network in multichain based NFTs marketplace provides users with greater flexibility and choice when it comes to transacting NFTs. By supporting multiple blockchain networks multichain NFTs marketplace allows users to access a wider range of NFTs and to choose the blockchain that best suits their needs in terms of speed, cost, and other factors (Christodoulou, et al., 2022). if a user wants to buy or sell an NFT quickly and at a low cost, they may choose to use a blockchain like Binance Smart Chain which typically has faster transaction times and lower fees than the Ethereum network. multichain NFTs marketplace include increased liquidity, greater accessibility, and more options for buyers and sellers. Multichain NFTs marketplaces also help to promote interoperability across different blockchain networks.

We have proposed research to design a multichain NFTs marketplace and its potential impact on the NFTs. The development of multichain NFTs marketplaces involves several technical challenges including interoperability, scalability, and security. The NFTs multichain based marketplace architecture is also discussed in paper that describes the bridge and cross communication concept which is allow for the transfer and communication of NFTs between different blockchain networks. Furthermore, discuss the layer architecture diagram can make it easier to understand the overall concept of the multichain NFTs marketplace architecture.

Ethereum is a primary mainchain blockchain that serves as the foundation for other blockchains called side chains. Sidechain is a separate blockchain that is attached to the Ethereum mainchain but operates with its own set of rules and protocols (Nejc & Diaci, Janez & Corn, & Marko, 2021). The Ethereum blockchain is responsible for validating all transactions that occur on the blockchain network (etherscan) (Wang, Mengjiao & Ichijo, Hikaru & Xiao, & Bob, 2020) and it is considered the most secure and reliable part of the blockchain.

The sidechains are used to perform specific functions that are not possible on the Ethereum blockchain. For example, sidechain may be used to carry out smart contract transactions, enable privacy features or facilitate faster and cheaper transactions. (Lee, Nam & Yang, , Jinhong & Onik, , & Mehed, 2019).

Multichain NFTs marketplaces also provide opportunities for investors and collectors to diversify their portfolios and reduce their exposure to a single blockchain. This thing can increase the demand and value of NFTs, as well as attract new users to the NFTs ecosystem. so, (Parrales, Gema & Batbayar, & Bayarc, 2022). However, the performance of

NFTs decides the direction of the future how many earn revenue from this market. It depend on market data in this paper we discuss the NFTs market passed data that show the NFTs value.

## II. RELETIVE WORK

Many researchers worked on NFTs based marketplace for more reliabilities, maintainability's, scalabilities, securities, simplicity and performance. Major focus of the researcher to enhanced the issue the singlechain NFTs marketplace and little bit work on the issue of interoperability, cross chain communication and multichain network for DApps in this section we discuss the related research work for NFTs and DApps. Here are some examples of research papers that have been written on it.

The author (Kireyev & Pavel, 2022) describe their paper excel in terms of timeliness, design considerations, market intelligence, and practical implementation potential. the paper addresses the growing trend of NFT marketplaces, it has gained significant prominence in the digital landscape. it gives attention to design aspects and provides practical guidance for developers and entrepreneurs seeking to create NFTs or improve marketplaces. However, its failure to address emerging technology advancement and protocols may refer to the impact of the design and functionality of NFT marketplaces. blockchain technology requires constant monitoring and analysis of new developments, such as layer solutions, scalability improvements, or advancements in token standards.

.The author (Behera, et al., 2023) aims to provide different strategies and features to enhance the user experience, token management, improve discoverability and promote transparency and trust within the marketplace ecosystem. The paper's purpose is to create an intuitive and accessible platform that attracts and engages a diverse user base that includes both novices and veterans of NFTs. However, it is fail to consider evolving user preferences and blockchain technological advancements which may result in an outdated design.

The author's (Do & Thuat, 2023) purpose of this research is to provide an in depth analysis of the evolution of blockchain technology and to propose a taxonomy for categorizing different generations of public blockchains. The taxonomy becomes more scientific and easier to apply uniformly across different blockchain systems by specifying particular criteria. however, there is no consensus within the blockchain community on how to classify public blockchains and a lack of coverage of a wide range of blockchain systems, which may lead to the omission of new or niche technology.

The author (Kadam, et al., 2023) provides a comprehensive in this paper. this paper provides an extensive literature review, illustrating a complete awareness of existing blockchain research, and covers an important and timely topic, discussing the construction of an NFT marketplace with various chains and digital currency exchange. The suggested approach provides an acceptable fix to the shortcomings of present NFT markets. However, The paper lacks a detailed evaluation of the proposed system, such as performance benchmarks, user feedback, or comparisons with existing NFT marketplaces. The suggested system's limitations and potential obstacles are not properly explained, leaving the opportunity for uncertainty and more research.

This paper (Das, et al., 2022) focuses on the NFT ecosystem's security challenges and market dynamics. It adresses to offer a thorough analysis of how the NFT ecosystem operates and to identify potential security vulnerabilities only for single chain marketplace that might result in financial losses and interactions between markets, external companies, and NFT ecosystem users. owever, It also examines possible weaknesses in the architecture of the top eight NFT markets by transaction volume and tackles the question of legitimacy in NFTs.

The purpose (White, Bryan & Mahanti, Aniket & Passi, & Kalpdr, 2022)of this study is to explore and analyse the NFTs industry, with a special focus on OpenSea, the largest NFTs marketplace.It states that NFTs are unique digital IDs that signify ownership of different cryptoassets, collectibles, and gaming assets on the Ethereum blockchain, and it emphasises the NFT market's quick growth in 2021 as well as the growing acceptance of blockchain and cryptocurrency technology.The study's findings demonstrate that a small subset of heavy users generates significant growth in the NFT sector.However, this paper fails to uncover the emergence of user communities within the network's sparsity, where most users tend to congregate, as well as its failure to address an in-depth study on NFT marketplaces other than opensea, which is based on a single chain the environment and a multi-level analysis of significant findings and trends.

The novelty of our work is that we have used multichain blockchain technology in NFT marketplace. We have solved the customer's trust issues like interoperability, scalability and flexibility by executing smart contracts for acceptance testing. We have also try solved single chain problem However, there are efforts underway to address this issue through the development of cross-chain bridges and interoperability protocols, which aim to enable NFTs to be transferred between different blockchain networks. Some examples of cross-chain solutions include Polkadot, Cosmos, and the Interledger Protocol. Despite these efforts, the issue of single chain compatibility remains a challenge for the wider adoption and mainstream use of NFTs.

## III. PRELIMINARIES

This chapter presents the preliminaries used in the NFTs multichain architecture. It describes the major components of multiple blockchain networks (e.g. Ethereum, Binance Smart Chain, Polygon, etc.) that have been used in architecture including Binance smart Chain, Smart Contracts, Wallet Integration & User Interface and Auctions and Sales,. Multichain Infrastructure.

### *A. ETHEREUM BLOCKCHAIN*

Ethereum is a decentralized, open source (Abichandani, Pramod, & Lobo, Deepan & Kabrawala, Smit & McIntyre, & William, 2020) blockchain with smart contract features. It is referred to as a public ledger since it is used to record and verify transactions. The Ethereum software platform contains its own digital currency, ether (ETH), which can be shared across accounts linked to Ethereum blockchain network. Users access their Ethereum accounts using Ethereum wallets or apps. This decentralized blockchain network also serves as a platform for DApps, smart contracts and data storage. It is disseminated by using a peer-to-peer (P2P) network (Farooq & Kalim,, Junaid & Rasheed, & Adnan, 2022).

### B. SMART CONTRACTS

Smart contracts are lines of code that are recorded on the Ethereum blockchain and automatically execute when specified terms and circumstances are met. These Ethereum smart contracts or transaction protocols are written in Solidity language and implemented on the blockchain to document occurrences in accordance with the terms of a protocol or contract (Andrea & Ibba, Simona & Baralla, Gavina & Tonelli, & Roberto & Marchesi, 2019). Smart contracts are used to limit the participation of third parties and harmful circumstances.

### C. Binance smart Chain

Binance Smart Chain (BSC) is a blockchain platform launched by Binance in 2020. It is designed to be compatible with the Ethereum virtual machine (EVM) (Monga & Suhasini & Dilbag, 2022), meaning that it can support smart contracts and DApps that were originally built for the Ethereum blockchain. BSC is built on a Proof of Stake (PoS) consensus mechanism, which makes it faster and cheaper to use than the Ethereum blockchain, making it a popular choice for developers who want to build decentralized applications with low transaction fees.

This can be especially important for NFTs based applications, where high transaction fees can be a major barrier to entry for both creators and buyers. BSC is the BEP20 standard, which is similar to the ERC20 standard used for tokens on the Ethereum blockchain (Cernera & La Morgia, Massimo & Mei, & Alessandro & Francesco, 2022). BEP20 tokens can be easily created and managed on BSC, making it easy for NFTs creators to tokenize their creations and sell them on a BSC based marketplace. Buyers can use Binance Coin (BNB) or other cryptocurrencies that are supported on the Binance Smart Chain to purchase these tokens. BSC's dual chain architecture allows for increased flexibility and interoperability between different blockchain networks, while its high throughput and low fees make it an attractive option for developers and users looking to create and use NFTs based DApps and marketplaces. BSC is built using the Tender mint consensus engine and Cosmos SDK, which allow for the creation and deployment of custom blockchain applications.

### D. Wallet

Users need to connect their wallets to the marketplace to interact with NFTs. The marketplace should support popular wallets like MetaMask, Wallet Connect, and Coinbase Wallet (Philipp & Lorimer, Anna Harbluk & Snyder, & Peter & Livshits, 2021). The marketplace should also have a built in wallet that users can use to hold their NFTs and other digital assets include Web3.js, Ethers.js, and the WalletConnect protocol (Jaynti & Nailwal, Sandeep & Arjun, & Anurag, 2021). The User interface should provide all the necessary information about NFTs, such as their ownership, history, and metadata. The marketplace should also provide users with tools for managing their NFT collections, such as the ability to create and manage collections, view transaction histories, and set alerts for new NFT listings. DApps are made up of a frontend (JavaScript, React, and Next.js) or user interface in order to communicate with the end users and a smart contract or backend code that operates on a decentralized P2P network.

### E. ERC 721

The ERC721 is NFTs Standard that implements an API for tokens within Smart Contracts (Shuo & Chen & Jiachi & Zheng,, 2023). ERC721 is the standard interface used for creating and managing NFTs on the Ethereum blockchain. NFTs creators can ensure that their NFTs can be easily traded and managed on a variety of different marketplaces and platforms, as long as those platforms also support the ERC721 standard. This standard specifies how NFTs should be produced, kept, and transferred, including the ability to authorize the transfer of a certain number of tokens from a particular account to another.

### IV. METRIAL AND METHODOLOGY

In this section we will discuss complete system designing. We break all solution in system component and explain all function of each component and elaborate all process of the system. Discuss the different entities of system and how to differentiate their process shows in figure 3 buyer, seller and artists interact with each other through NFTs marketplace all entities call functions on the multichain blockchain network. Important is that how the communicate different blockchain network with each other. According to figure 3 all participant entities must be register on the system for authentication, verification, and accountabilities. Artists create NFTs and list them on the marketplace to sell their contributions to interested buyers. The Buyer can buy the NFTs across the network and similarly, artist can sell their NFTs across the blockchain network like binance, polygn and Ethereum. The most of NFTs project are the off chain type and they are usually represented by a unique token or hash value stored with smart contract which is associated link with external storage "IPFS" where image and metadate itself stored.

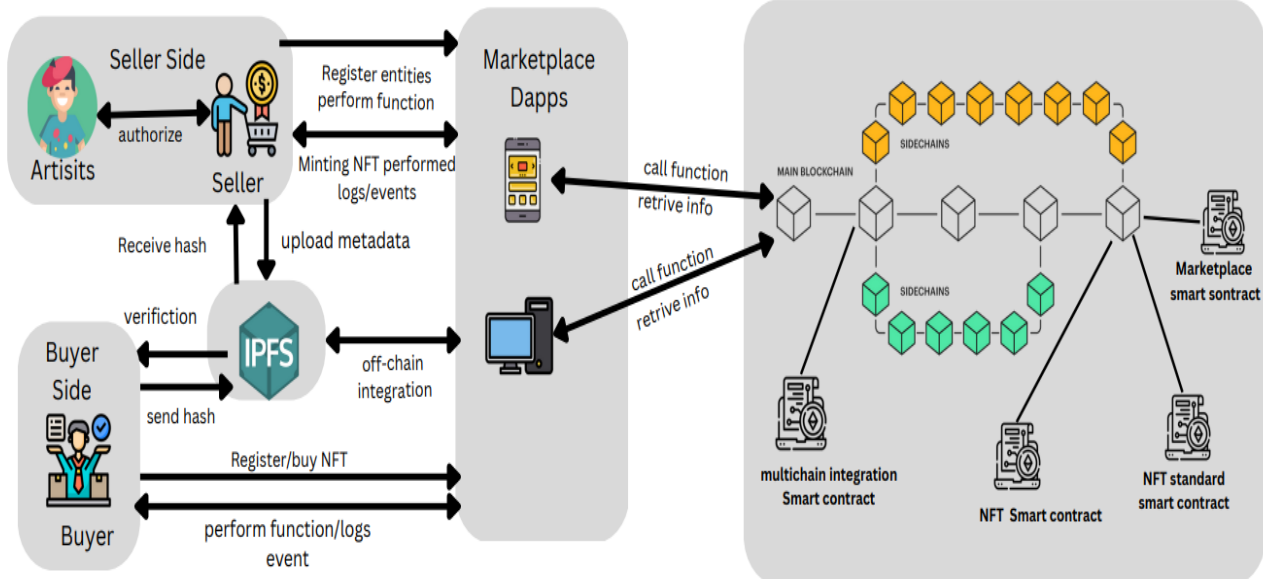

**Figure 3** *system architecture*

### A. Role of entities
Now we discuss all entities of the NFTs marketplace. artists create NFTs for and list down on the marketplace. In the multichain NFTs marketplace the seller plays a crucial role in facilitating the buying and selling of NFTs across different blockchain networks. Seller or artist responsible to create and selling NFTs on the marketplace including providing all detail such as price, attribute and its description. According to NFTs architecture seller responsible to managing NFTs such as updating their metadata and transfer the ownership to the buyer when the transaction is complete. Seller needs to familiar with different blockchain and adapt their NFTs to work seamlessly across them seller. Another entity buyer plays important role in NFTs architecure when the buyer purchase NFT from the creator or seller. Buyer has power to drive demand and increase the value of NFTs, which can encourage the creator to produced high quality content. Buyer maybe involves to governing decision that can directly affect the functionality of platform and tokens economy. Buyer can hold and trade to potentially profit from increasing the value of NFTs.

All participants have must crypto wallet addresses, in the marketplace every entities to perform the action like buy or sell NFTs must hold crypto wallet addresses, when entities registered on on chain network they can trackable on the network. participant must be on chain and off register, for off chain registration participant must be following KYC verification procedure to ensure the proof of identities. NFTs multichain marketplace based on proof of stake means that ensure the seller is trustworthy and honest. Proof of Stake is a consensus process used in blockchain-based networks to confirm transactions and produce new blocks, but it has nothing to do with how the NFTs marketplace operates.It was utilized to reward participants.

### B. Multichain
Basically multichain technology can be use to enable to creation and transfer of NFT to different blockchain network and it allow the more flexibility and interoperability .to implement the NFTs multichain architecture a single blockchain network the support NFTs then create subchain or sidechain to enable to transfer NFTs to different networks. For example, ERC 721 and ERC 1155 standard used for NFTs in Ethereum network .this chain interconnect with sidechine (such as Polygon and Binance Smart Chain) is a common design pattern in blockchain systems that allows for interoperability between different blockchains. The second approach a multichain network is to create a standard protocol that can be implemented across multiple chains. This can be achieved by using interpretability or cross chain bridges which is allow for the transfer of assets or tokens between different blockchain networks.

### C. Market smart concept
We have two smart contracts in our solution that we generated beyond of the ERC721 standard. The ERC721 OpenZeppelin protocol using their smart contract libraries, the smart contracts are utilized to produce ERC721 tokens. The marketplace smart contracts are primarily in charge of registering the companies that participate and managing the buying and selling of NFTs. To mint or trade NFTs, all associating and participating entities on the chain must be authorized through the marketplace. Enrollment is only complete if users submit off-chain evidence of identification via formal certified papers such as copies of their passports or valid identification cards.

### D. Seller smart contract

A registered seller owns the NFTs seller smart contract. To mint their own NFTs, each authorized seller will have their own NFTs seller smart contract. The marketplace smart contract receives the address of the NFTs seller smart contracts. NFTs seller smart contracts generate NFTs by using NFTs metadata like as the URI, price, IPFS hash, and ID. Furthermore, if all circumstances are satisfied, this contract gets calls from the marketplace smart contracts for transferring the ownership of a token to a buyer. The functions and properties of two ERC721 SCs are passed down to the NFTs seller smart contracts. The inherited functions and characteristics are from the ERC721URIStorage SC, which in turn inherits the ERC721 smart contract's functions and properties.

### E. Buyer price fix

An NFT listed on the market might be sold as a fixed price NFT or via an auction. The seller has complete control over how the NFT is sold. The process for a fixed-price NFT begins with an interested customer wanting to purchase the NFT through the marketplace. The buyer asks the marketplace for the NFT. The blue arrows represent an off-chain interaction for this request. The marketplace then links to the blockchain via the marketplace SC, as seen by the brown and red arrows, indicating an on-chain interaction. If all requirements are satisfied in the marketplace, SC sends the Ether or the asking rate of the NFT to the seller. For example, both the buyer and seller must be enrolled, and the seller's NFT SC address must be in the marketplace SC's preserved list of NFT seller SCs. Furthermore, the price paid by the buyer must be the same as the price of the NFT requested by the seller. If all requirements are satisfied, the marketplace SC transfers the price to the seller's account and requests that the NFT seller SC transfer the buyer's ownership of the NFT. The ownership is subsequently transferred through the ERC721 SC by the NFT selling SC, and the new ownership is declared. A buyer can then use the IPFS hash to validate the NFT. Ratings can also be submitted by both the purchaser and the vendor.

### F. Buying auction

NFTs can be purchased at auction. The auction is conducted in the NFT marketplace, and the beginning price is set by the seller. The seller starts the auction, and the buyers begin bidding before the sale ends. The decentralized application timer is used to timing the auction. Figure 3 shows the DApp triggering a call to the marketplace and the auction ending. When a buyer offers, the bid is sent to the marketplace SC. Once a bid is placed, it cannot be changed. When the auction concludes, the marketplace SC requests a transfer of ownership from the NFT seller SC. The vendor transfers ownership of the NFT to the highest bidder. The remaining bidders have their bids refunded. This strategy assures that every bidder is a serious buyer and that no collateral is required to bid. Other auctions, in which bidders just make bids without transferring Ethers, run the possibility of buyers changing their minds or bidding large sums only to withdraw, allowing another bidder with a cheaper offer to win the auction. As a result, in this scenario, consumers must pay an assurance before making an offer, and the collateral is not repaid if a bidder is dishonest. In addition to decreasing their reputation, the collateral placed aids in the eradication of dishonest buyer activity. We encourage the technique in which bidders pay the whole amount of their bid when they submit their offer. The approach selected guarantees that the total bids put are large enough to deter dishonesty and eliminate illegal behavior.
.

### Layer architecture

The now discuss architecture layer of the NFT marketplace shows in the figure.

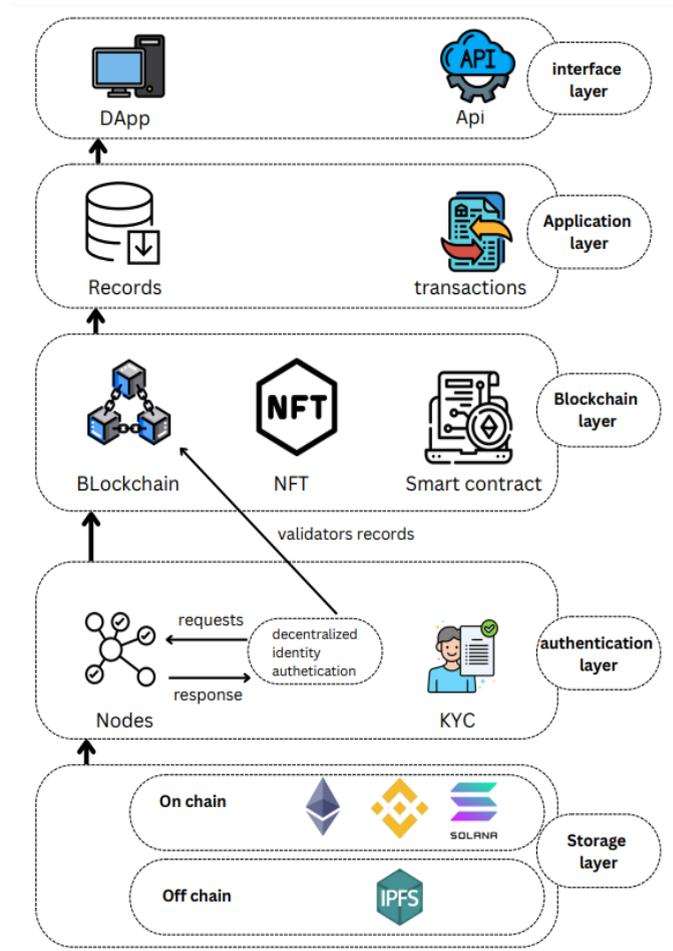

1. **UI Layer**

A user interface in the multichain NFT marketplace UI layer allows users to purchase, sell, and trade NFTs across different blockchain networks. A user-friendly interface that allows consumers to simply browse around the platform and discover NFTs that they are interested in purchasing is often included in the UI layer. It also includes tools to help customers construct, list, and control their own NFTs for sale.

2. **Application Layer**

In the multichain NFT marketplace application layer refers to a platform that enables users to buy, sell, and trade NFTs (non-fungible tokens) across different blockchain networks. The multichain NFT marketplace application layer enables users to interact with different blockchains, including Ethereum, Binance Smart Chain, Polygon, and others

3. **Blockchain Layer**

In a multichain NFT marketplace, the blockchain layer plays a critical role in ensuring the security, transparency, and immutability of NFT transactions. The blockchain layer of a multichain NFT marketplace presents several benefits, such as increased security, immutability, and transparency of NFT transactions. However, it also poses several challenges, such as the need for interoperability between different blockchain networks, scalability concerns, and high transaction costs.

4. **Authentication Layer**

In the context of a multichain NFT marketplace, the authentication layer would need to be designed to support multiple blockchains and ensure that users can access and trade NFTs across different networks without compromising security or trust. this layer using KYC processes on a blockchain involves creating a decentralized system that allows users to securely and privately verify their identities. .This can involve integrating with various blockchain networks and protocols, designing secure and user-friendly interfaces, and implementing robust security measures to protect user data and prevent fraud or hacking.

5. **Storage Layer**

In a multichain NFT marketplace, NFT data needs to be stored and managed across different blockchain networks, which can vary in terms of their storage capacity, security features, and accessibility. The storage layer needs to ensure that NFT data is properly encrypted, replicated, and backed up to prevent data loss, corruption, or unauthorized access.

## Multichain Architecture Protocol

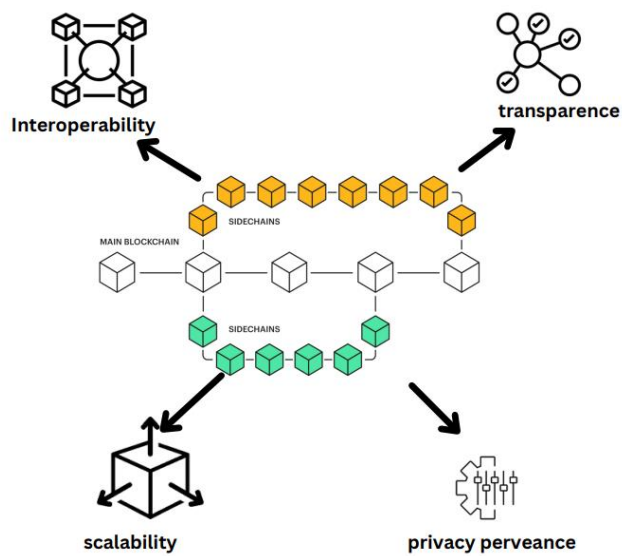

Already discuss the multichain concept know in this section discuss the details multichain architecture protocol, in this architecture, each blockchain network is designed to serve a specific purpose or function, and they are interconnected to enable seamless communication and data transfer between them. The multichain architecture is particularly useful in situations where a single blockchain network cannot accommodate all the necessary functionalities or handle the required transaction volume. By having multiple blockchain networks working in tandem, the architecture can leverage the unique strengths of each blockchain network to provide a more comprehensive solution. Each blockchain network in the architecture can be optimized for a specific use case or function, such as fast transaction processing, complex smart contract execution, or data storage. Interconnected blockchain networks can communicate with each other using various protocols and standards. This enables the exchange of value and data between different blockchain networks, which enhances the overall interoperability and flexibility of the system. According multichain architecture protocol is a set of rules and standards that govern the communication and interoperability between multiple blockchain networks in a multichain architecture. protocol defines how different blockchain networks can communicate with each other and how they can exchange value and data. so the protocol in a multichain architecture is designed to enable seamless communication and data transfer between different blockchain networks. This involves creating a common set of standards and protocols that can be used by all the blockchain networks in the architecture.

1. **Cross-chain communication**

The key feature of multichain protocols, enabling different blockchain networks to communicate and interact with each other. In a multichain architecture, there are multiple independent blockchain networks that operate in parallel, and cross-chain communication is necessary to enable the exchange of value and data between them. There are different methods of cross-chain communication, and multichain protocols define a set of standards and protocols for this communication. One common method is through the use of bridges, which are smart contracts or other intermediary components that facilitate communication between different blockchain networks. Bridges enable the transfer of assets or data from one blockchain network to another by locking the assets or data in one network and creating a corresponding asset or data representation on another network. This allows the recipient network to access and use the transferred assets or data, even though they are not natively stored on that network. Another method of cross-chain communication is through the use of interoperability protocols, which enable different blockchain networks to communicate directly with each other. These protocols define a common set of standards and protocols that all participating networks must adhere to, enabling them to exchange information and value seamlessly.

2. **Interoperability**

Interoperability is a critical feature of multichain protocols that enables different blockchain networks to communicate and work together seamlessly. Interoperability protocols provide a set of standards and protocols that allow different blockchain networks to interact with each other, exchange value and data, and execute smart contracts across different networks.There are various approaches to achieving interoperability in multichain protocols. One common approach is to use a common standard, such as the Ethereum Virtual Machine (EVM), which allows different blockchain networks to execute smart contracts using the same programming language and virtual machine. This allows smart contracts to be executed seamlessly across different blockchain networks, even if they have different consensus mechanisms or native tokens. Another approach to interoperability is through the use of cross-chain bridges. Bridges are smart contracts that allow assets or data to be transferred from one blockchain network to another, creating a corresponding asset or data representation on the recipient network. This enables the recipient network to access and use the transferred assets or data, even if they are not natively stored on that network.

3. **Consensus mechanism**

Multichain protocols use different consensus mechanisms to achieve agreement among the network participants on the state of the blockchain ledger. Consensus mechanisms are critical in ensuring the integrity and security of the blockchain network, and they determine how transactions are verified and added to the blockchain. There are different types of consensus mechanisms used in multichain protocols, every mechanism have its own strengths and weaknesses.like proof of work , proof of stack, Byzantine Fault Tolerance and Proof of Authority. There are many other consensus mechanisms used in multichain protocols, and each has its own strengths and weaknesses. The choice of consensus mechanism depends on the specific requirements of the blockchain network, such as its security, scalability, and decentralization goals.

***Sidechain***

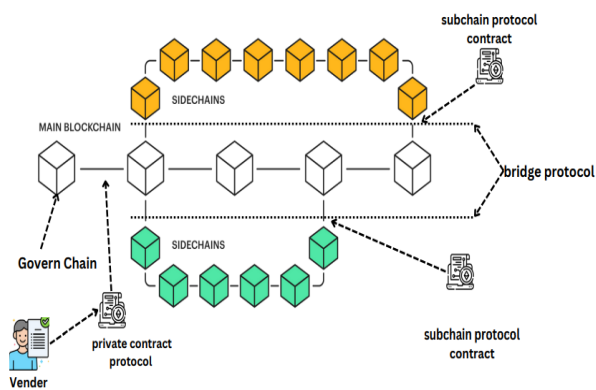

Sidechain architecture is a design pattern for blockchain networks that allows for the creation of separate chains that are interoperable with the main blockchain network. Sidechains are designed to enable the creation of new applications, services, and assets that can leverage the security and immutability of the main blockchain network, while also providing a higher degree of flexibility and scalability. In a sidechain architecture, the main blockchain network acts as the anchor chain, while the sidechains are attached to the anchor chain as independent, parallel chains. The sidechains operate independently from the main chain, but they can still communicate with the main chain using a two way peg system, which enables the transfer of assets and data between the two chains. The two-way peg system works by allowing users to lock up a certain amount of cryptocurrency on the main chain, which is then converted to a corresponding number of tokens on the sidechain. The user can then use the tokens on the sidechain to transact with other users on that chain. When the user wants to move the tokens back to the main chain, they must first burn the tokens on the sidechain, which releases the corresponding amount of cryptocurrency on the main chain. One of the benefits of sidechain architecture is that it enables the creation of specialized chains that can handle specific use cases, such as high-speed transactions, privacy-focused transactions, or complex smart contracts. This can help to improve the overall scalability and flexibility of the blockchain network, while also providing a more tailored experience for users. Another benefit of sidechain architecture is that it can help to reduce congestion on the main chain, by offloading some of the processing and transactional load to the sidechains. This can help to reduce transaction fees and improve transaction throughput on the main chain, while also providing a more efficient and streamlined user experience.

In a NFT marketplace sidechain architecture, the main blockchain network acts as the anchor chain, while the sidechain is attached to the main chain as an independent, parallel chain. The sidechain is designed specifically to handle NFT marketplace transactions and can be customized to include features such as increased transaction speed, reduced fees, and improved NFT management tools. The two-way peg system used in NFT marketplace sidechain architecture is similar to the system used in other sidechain architectures. Users can lock up a certain amount of cryptocurrency on the main chain, which is then converted to a corresponding number of tokens on the NFT marketplace sidechain. These tokens can then be used to buy, sell, and trade NFTs on the sidechain. The benefits of NFT marketplace sidechain architecture include improved transaction speed and reduced fees, which can help to make NFT marketplace transactions more accessible to a wider range of users. It can also provide a more efficient and streamlined user experience, with specialized tools and interfaces for buying, selling, and trading NFTs.

### V. implementation

In this section, we will discuss the implementation details and algorithm of the smart contract, and we will test implemented algorithms. The smart contracts written in solidity language using the Hardhat and wallfle. Hardhat and waffle is main two popular developing platform in which allows the writing, developing and testing the smart contract. To facilitate the minting process of NFTs, we utilize the OpenZeppelin library's built-in capabilities with Hardhat in combination with Waffle. There is three main smart contract named marketplace, seller NFT and multichain smart contracts. NFT smart contract connects the code to OpenZeppelin library by inheriting the ERC721URI storage. ERC721URIStorage is a smart contract from the OpenZeppelin library that provides storage and functions related to the of ERC721URI tokens. Therefore, in the implementation of NFT contract on multichain each blockchain have own specification and requirements for handling tokens metadata and URI.in additionally we add two functions for flexibility and control multichain NFT contract. For instance, when minting the NFT generate unique IPFS hash NFT metadata. Therefore, we tied the NFT identification (ID) to a unique hash and pricing value. When selling an NFT in an auction, the stored price serves as the NFT's starting price. Figure 4 depicts the Marketplace smart contract and the NFT Seller smart contract, both of which were constructed using our code and have all of their functionalities and features. It also displays the two SCs inherited by the NFT Seller SC from the ERC721 OpenZeppelin library. In the graphic, we have just presented the features and functionalities that we have added to the current ERC721 OpenZeppelin SCs. The diagram depicts inheritance using blue arrows. As a result, the NFT Seller SC is an ERC721URIStorage SC, and so an ERC721 SC.And various artists, buyers, and sellers are registered on the market place SC.

### A. *Brigde Contract*

**Algorithm 1** Bridge Contract
1: **Contract Variables:**
2: *ethBridge:* Ethereum bridge contract address
3: *bscBridge:* Binance Smart Chain bridge contract address
4:
5: **Event:**
6: $NFTLocked(tokenId, fromChain, toChain, nftContract)$ : NFT lock event
7: $NFTUnlocked(tokenId, fromChain, toChain, nftContract)$ : NFT unlock event
8:
9: **Function lockNFT(nftContract, tokenId):**
10:    Require that the caller is the owner of the NFT
11:    Transfer the NFT from the caller to the bridge contract
12:    Emit $NFTLocked(tokenId, fromChain, toChain, nftContract)$ event
13:
14: **Function unlockNFT(tokenId, nftContract):**
15:    Require that the caller is the Binance Smart Chain bridge contract
16:    Transfer the NFT from the bridge contract to the caller
17:    Emit $NFTUnlocked(tokenId, fromChain, toChain, nftContract)$ event
18:
19: **Function setBridges(ethBridgeAddress, bscBridgeAddress):**
20:    Require that the bridge addresses are not already set and are valid addresses
21:    Set *ethBridge* to *ethBridgeAddress*
22:    Set *bscBridge* to *bscBridgeAddress*

This algorithm performs as a bridge contract, allowing NFTs to be locked and unlocked between Ethereum and Binance Smart Chain.The "lockNFT" method moves an NFT from the caller's account to the bridge contract, whereas the "unlockNFT" function returns the NFT to the caller's account on the original chain. The "setBridges" function is used to configure the bridge contracts' addresses.The method generates events to monitor the locking and unlocking of NFTs between chains.The algorithm defines two contract variables, "ethBridge" and "bscBridge," that respectively store the addresses of the Ethereum bridge contract and the BSC bridge contract. These variables are used to locate the associated bridge contracts.The algorithm distinguishes two events: "NFTLocked" and "NFTUnlocked." When an NFT is locked (transferred) from one chain to another, the "NFTLocked" event is emitted, and it contains the token ID, the origin chain, the target chain, and the NFT contract address.When an NFT is released or transferred back to its originating chain, the "NFTUnlocked" event is emitted, and it has the same parameters as the "NFTLocked" event. In function lockNFT(nftContract, tokenId) The owner of an NFT can use this function to lock (transfer) it from their account to the bridge contract. It accepts as inputs the NFT contract address and the token ID. Before proceeding with the transfer, the algorithm checks that the caller (transaction originator) is the actual owner of the NFT. NFT is moved from the caller's account to the bridge contract if the caller is the owner. Following the transfer, the algorithm broadcasts the "NFTLocked" event, sending as arguments the token ID, the source chain (Ethereum), the destination chain (BSC), and the NFT contract address. This event indicates that the NFT has been locked and is now available for unlocking on the opposite chain. Function unlockNFT(tokenId, nftContract): The bSc bridge contract may use this function to unlock (transfer) an NFT back to its native chain. It accepts as arguments the token ID and the NFT contract address.Before proceeding with the unlock, the algorithm confirms that the caller is the Binance Smart Chain bridge contract. If the caller is the proper bridge contract, the NFT is sent back to the caller's account on the original chain from the bridge contract.

Following the unlock, the algorithm broadcasts the "NFTUnlocked" event, sending as arguments the token ID, the source chain (BSC) the destination chain (Ethereum), and the NFT contract address. This occurrence signifies that the NFT has been unlocked and is now accessible via its original chain. Function setBridges(ethBridgeAddress, bscBridgeAddress): Sets the addresses of the Ethereum bridge contract and the BSC bridge contract. The method ensures that the bridge addresses are not already in use and are legitimate. If the addresses are correct, the algorithm changes the "ethBridge" variable to the Ethereum bridge contract address and the "bscBridge" variable to the BSC bridge contract address.

### B. Escrow contract

```
Algorithm 1 Escrow Contract
1: Contract Variables:
2:   buyer: Address of the buyer
3:   seller: Address of the seller
4:   arbitrator: Address of the arbitrator
5:   amount: Amount of funds involved in the escrow
6:   state: Current state of the escrow (AWAITING_PAYMENT, AWAIT-
       ING_DELIVERY, COMPLETE, REFUNDED)
7:
8: Constructor:
9:   Initialize buyer, seller, arbitrator, amount, and set state to AWAIT-
       ING_PAYMENT
10:
11: Function confirmPayment():
12:   Require that the caller is the buyer
13:   Require that the sent value is equal to amount
14:   Require that the state is AWAITING_PAYMENT
15:   Update state to AWAITING_DELIVERY
16:
17: Function confirmDelivery():
18:   Require that the caller is either the buyer or the seller
19:   Require that the state is AWAITING_DELIVERY
20:   if caller is buyer then
21:     Update state to COMPLETE and transfer amount to the seller
22:   else
23:     Update state to REFUNDED and transfer amount to the buyer
24:   end if
25:
26: Function refund():
27:   Require that the caller is the arbitrator
28:   Require that the state is AWAITING_DELIVERY
29:   Update state to REFUNDED and transfer amount to the buyer
```

The purpose of an escrow contract is to act as a trusted intermediary between the buyer, seller, and arbitrator, to providing a secure mechanism for holding funds during a transaction until certain conditions are met. The contract includes several variables buyer, Seller ,arbitrator and amount shows in algorithm. State variable is used to keep track of the current state of the escrow contract in which include various states such as AWAITING_PAYMENT, AWAITING_DELIVERY, COMPLETE, and REFUNDED. state variable is essential for managing the flow of the transaction and ensuring the contract follows to appropriate actions and conditions based on its current state.The contract's constructor is executed when the contract is deployed and initializes the contract variables with the provided inputs variables and It also sets the initial state to AWAITING_PAYMENT. Function confirmPayment() called by the buyer to confirm the payment and initiate the escrow. It is verify that the caller is the buyer, the sent value matches the specified amount, and the state is AWAITING_PAYMENT. If all conditions are met, the state is updated to AWAITING_DELIVERY. buyer or the seller can call the function confirmDelivery() to confirm the delivery of the funds. the caller is the buyer, the state is updated to COMPLETE, and the funds are transferred to the seller. If the caller is the seller, the state is updated to REFUNDED, and the funds are returned to the buyer. The function refund() can only called by arbitrator to initiate a refund in case of disputes or unresolved issues. It checks that the caller is the arbitrator and that the state is AWAITING_DELIVERY. If both conditions are met, the state is updated to REFUNDED, and the funds are returned to the buyer.

### C. Token contract

```
Algorithm 1 Token Contract
1: Contract Variables:
2:   balances: Mapping of account addresses to token balances
3:
4: Event:
5:   Transfer(from, to, amount) : Token transfer event
6:
7: Function balanceOf(account):
8:   Return the balance of the specified account
9:
10: Function transfer(to, amount):
11:   Require that the sender has a sufficient balance
12:   Reduce the sender's balance by the transfer amount
13:   Increase the recipient's balance by the transfer amount
14:   Emit Transfer(msg.sender, to, amount) event
```

This smart contract allows essential functionalities for token transfers and balance tracking. The contract enables the creation of a specific token by defining its properties, such as name, symbol, and decimal places. It determines the initial supply of tokens and assigns them to the contract deploys account. It maintains a mapping of account addresses to their corresponding token balances. This allows any account holder to query their balance at any given time. The balance tracking feature ensures accurate and transparent token ownership. Token contracts enable the transfer of tokens between different accounts. Account holders can initiate transfers by specifying the recipient's address and the desired amount. It's verifying that the sender has sufficient balance and updates the balances accordingly. It emits a transfer event to notify external systems about the transaction.

### D. Markpace smart contracts

The marketplace smart contract is a more complex because it's doing a lot more custom functions that are not already implemented in OpenZeppelin.the contract allows to user to basic functionality such as List an item for sale, Sell an item, Buy an item and Unlist the item. Each item listed in the marketplace is represented by a struct Item, which includes a ID, the seller's address, price, and boolean flag to indicate if the item is sold. We describe one by one.

## 1) List NFT

```
Algorithm 1 NFT Listing
1: nftContract: Address of the NFT contract
2: tokenId: Token ID of the NFT to be listed
3: price: Listing price of the NFT
4: chain: Blockchain chain on which the NFT is being listed
5:
6: Function listNFT():
7: Require: Valid inputs and caller is the NFT owner
8: Require: NFT contract supports the specified chain
9: Increment listingCount and create a new Listing with the given details
10: Emit NFTListed event
```

NFT listing part of marketplace algorithm checks the validity of the inputs and ensures that the caller is the rightful owner of the NFT and It also verifies whether the NFT contract supports the specified blockchain chain to ensure compatibility. When the inputs are validated, this part of algorithm proceeds to increment the listing count and creates a new Listing. NFT listing part typically includes information such as the seller's address, the NFT contract address, the token ID, the listing price, and the listing status. emits NFTListed event to notify interested parties about the successful listing of the NFT. This event serves as a signal for potential buyers and facilitates the transparent functioning of the marketplace.

## 2) NFT Selling

```
Algorithm 1 NFT Selling Algorithm
1: nftContract: Address of the NFT contract
2: tokenId: Token ID of the NFT to be sold
3: price: Selling price of the NFT
4:
5: Function sellNFT():
6: Require: Valid inputs and caller is the NFT owner
7: Require: NFT is not already listed for sale
8: Approve the marketplace contract to transfer the NFT on behalf of the seller
9: Create a new SellingOrder with the given details
10: Emit NFTListedForSale event
```

SellNFT function that takes the NFT contract address, the token ID of the NFT to be sold, and the desired selling price as inputs. It checks the validity of the inputs and ensures that the caller is the owner of the NFT. It also verifies that the NFT is not already listed for sale in the marketplace. after successful validation the algorithm approves the marketplace contract to transfer the NFT on behalf of the seller. then creates a new SellingOrder in which typically contains information such as the seller's address, NFT contract address, token ID, selling price, and order status.At last, the algorithm emits an NFTListedForSale event to notify interested parties that the NFT is now listed for sale in the marketplace. NFTsell provides a standardized and secure approach to selling NFTs in a marketplace, ensuring transparency and facilitating smooth execution of NFT transactions.

## 3) Buying NFT

```
Algorithm 1 NFT Buying Algorithm
1: listingId: ID of the listing to purchase
2:
3: Function buyNFT():
4: Require: Valid listing ID and sent value is sufficient
5: Get the listing details based on listingId
6: Require: Listing is active and NFT ownership hasn't changed
7: Transfer the NFT from the seller to the buyer
8: Update listing status to inactive
9: Transfer payment to the seller
10: Emit NFTBought event
```

According to algorithm assumes the existence of a buyNFT() function that takes the ID of the listing to buy as input. first verifies the validity of the listing ID and ensures that sent value is sufficient to complete the purchase.the buying function retrieves the listing details based on the given listing ID and checks if the listing is active and if the ownership of the NFT hasn't changed when the listing was created. According to algorithm transfers the NFT from the seller to the buyer and updating the listing status to inactive to prevent further purchases of same NFT. It also transfers the payment from the buyer to the seller to complete the transaction. emits dsuccessful purchase of the NFT.

## 4) Unlist NFT

```
Algorithm 1 NFT Unlisting Algorithm
1: listingId: ID of the listing to unlist
2:
3: Function unlistNFT():
4: Require: Valid listing ID and caller is the listing owner
5: Get the listing details based on listingId
6: Require: Listing is active
7: Update the listing status to inactive
8: Emit NFTUnlisted event
```

This algorithm figure shows unlistNFT() function in which takes the ID of the listing to unlist as input. first verifies the validity of the listing ID and ensures that the caller is the owner of the listing.after successful validation to retrieves the listing information based on the provided listing ID to checks if the listing is currently active. Next step updates the listing status to inactive and effectively unlisting the NFT from the marketplace. This prevents further purchases of the NFT through the marketplace.Finally emits an NFTUnlisted event to notify interested parties about the successful unlisting of the NFT.

## VI. Performance and testing

In this section we go through over the solution. There are three parts of the solution: first upload and NFTs using different blockchain networks and also check bridge connection.

### E. Dashbroad

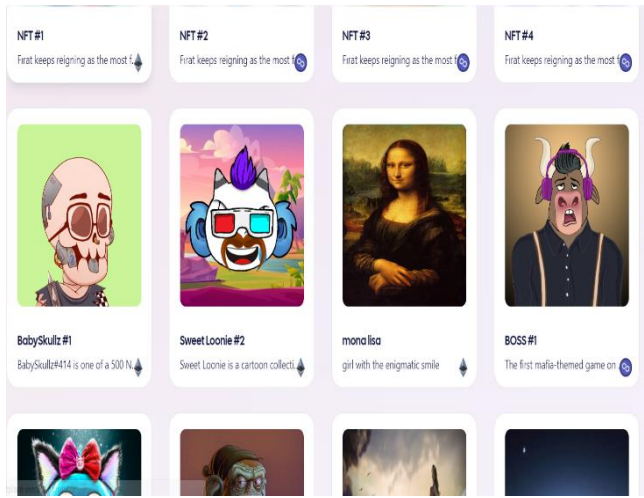

Moriarty NFTs have been created and traded on the ETH blockchain which has primary blockchain for NFTs development and activity. However, the multichain NFTs gallery provides a unified platform in which users can access NFTs from different blockchain networks, allowing for greater flexibility and accessibility in the NFT market. These platforms support NFTs from blockchains such as Ethereum Polygon, and others, enabling users to explore and trade NFTs across different architectures. You can see in the figure the Ethereum and polygon based NFT use in the same dashboard. Users can buy different same platforms.

### F. Bridge connection

for moving your NFTs from one blockchain to another, you can utilize bridge solutions that allow for cross chain transfers. These bridge solutions are designed to facilitate the transfer of NFTs, between different networks.

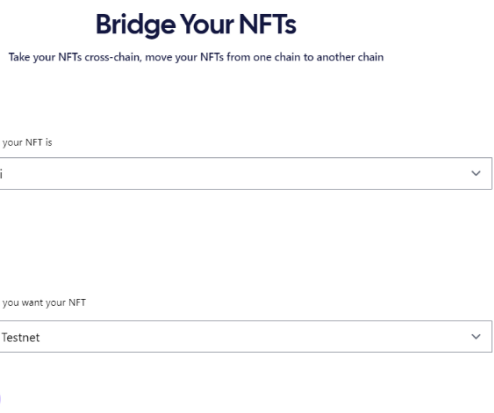

Look for a bridge solution that supports both the source and destination blockchains involved in the transfer. bridge platforms include swap, Polygon network, and Ethereum, Connect your wallets to both the source and destination blockchains. This typically involves using browser extensions like MetaMask. Follow the instructions provided by the bridge solution to initiate the transfer. This usually involves specifying the NFTs you want to transfer, source chain, destination chain and other relevant details.

### G. Transaction track

This is the wallet that deployed the smart contract and receives the 2.5% commission on each sale. This is the wallet that creates the NFT and lists it for sale This is the wallet that buys the NFT Use a faucet as shown earlier or send eth to the additional accounts you need to fund them with enough money to perform the transactions. Then you can click on the token in the block explorer. Here is a token that went through all these steps. You can also see this token's transactions in the block explorer .

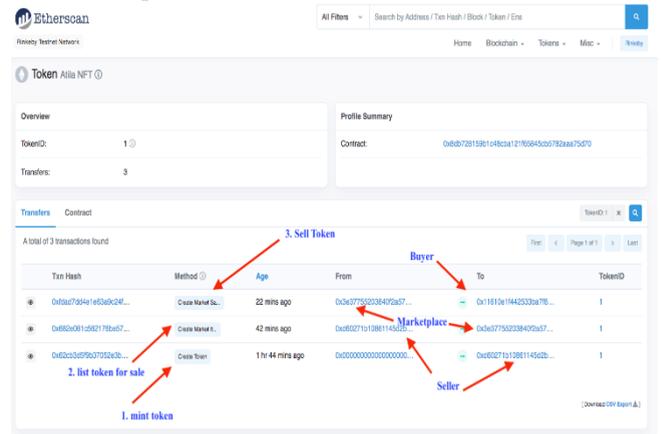

## DISCUSSION AND CONCLUSION

In this section, we discuss the NFT market & cap and trading volume of NFTs that make them popular. Also, and discuss the future of NFT.

### H. Market Cap

In this case, NFTs grew substantially in 2021, but this rise was inconsistent and has since levelled out in 2022. We'll look at how the NFT market has developed and how the market has fluctuated between 21-22.

NFT transaction volume has progressively increased since the beginning of 2021, however this rise varies. In 2022, NFTs activity ebbs and flows month to month; the value transmitted to NFTs markets increased in January, fell in February, and then began to rise in mid-April. The NFTs market went through a period of intense hype and popularity during that time, with high profile sales, celebrity endorsements, and widespread media attention. However, like any emerging market, NFTs space has experienced fluctuations and periods of contraction. Following the peak of the NFT boom in early 2021, there was a subsequent cooling off period where the market experienced some correction and consolidation. This leveling off in activity and market sentiment was expected as the initial hype subsided, and participants evaluated the long term potential and sustainability of the NFTss ecosystem.

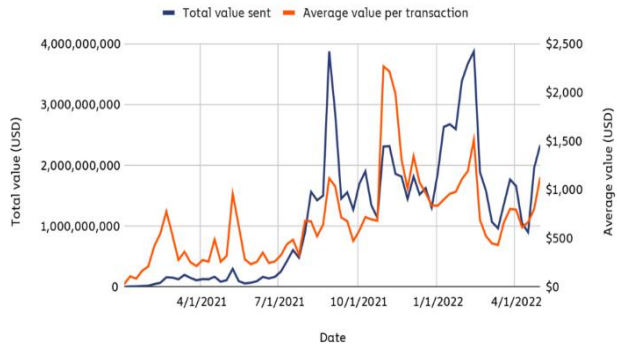

As of the beginning of May, collectors had transferred over $37 billion to NFTs markets overall in 2022 which puts it on track to exceed the $40 billion sent overall in 2021. The release of the Mutant Ape Yacht Club collection may have contributed to the late August spike in activity, and the LooksRare NFTs marketplace's debut in late January or early February of 2022 may have contributed to the late January to early February 2022 spike in activity. However, since late summer 2021, NFTs transaction growth has occurred in fits and starts, with activity largely remaining flat.

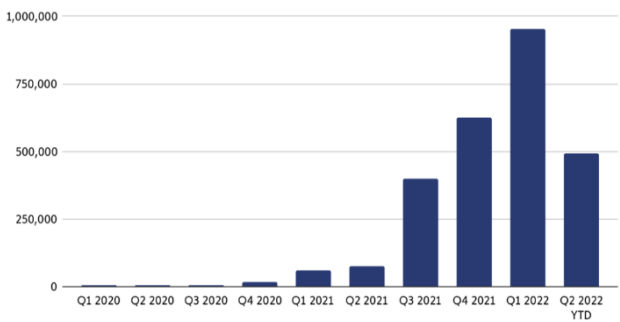

In the first quarter of 2022, 950,000 unique addresses purchased or sold NFTs, up from 627,000 in the fourth quarter of 2021. Since Q2 2020, the number of active NFTs buyers and sellers has risen in each quarter. As of May 1, 2019, 491,000 addresses have transacted with NFTs in Q2 2022, putting the NFTs market on track to maintain its quarterly growth pattern in terms of participants.

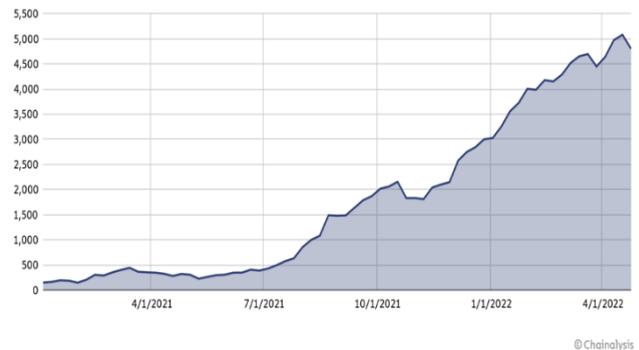

The number of active NFT collections on OpenSea, defined as those with any transaction activity in a given week, has likewise increased steadily since March 2021, and now exceeds 4,000.

### THE NFTs FUTURE AND WEB 3.0

NFTs play an important role in the development of Web 3.0 for the next generation of the Internet. NFTs enable real ownership of digital assets by leveraging blockchain. In Web 3.0 NFTs can be used to represent different digital assets, including virtual projects, art, virtual goods, and more. These tokens provide verifiable proof of ownership, scarcity and transferability giving individuals direct control in a decentralized manner. Web 3.0 aims to control and empower users. NFTs align with this vision by removing intermediaries and enabling peer-to-peer transactions. In Web 3.0, users can participate in the trading of NFT markets, create and interact with digital assets directly without relying on a central platform. This change promotes greater inclusion, economic empowerment and creative freedom for individuals.

### VALUE ADD DROPS

Value added drops (VADs) have the potential to shape the future of NFTs by additional benefits and experiences for collectors and NFTs enthusiasts. VADs are a concept where NFTs owners receive additional rewards or access based on ownership of a specific NFT. here are some impacts of VADs.

### 1. Enhanced Value and Scarcity

VADs can add value to NFTs ownership by providing unique benefits that are specific to certain NFTs holders. These benefits may include airdrops of coming NFTs, exclusive access to events or experiences, early access to new releases, utility tokens and other forms of digital rewards. By attaching these additional incentives to specific NFTs, VADs increase the perceived value and scarcity of these tokens, making them more desirable among collectors.

2. Marketing and promotion

VADs can be used as a marketing strategy to generate buzz and make interest in particular NFTs releases. By teasing additional rewards or benefits connected with owning certain NFTs creators and platforms can raise expectations and drive demand. This marketing method can help attract new collectors and create a sense of urgency among existing ones, increasing sales and activity in the NFTs marketplace.

### 3. Cross Promotion and Partnerships
VADs can facilitate collaboration and cross promotion between different projects and creators with in the NFTs space. For example, NFTs artists can collaborate with a musician and by owning their respective NFTs, collectors can gain access to exclusive music tracks or concert tickets. These partnerships can expand the reach and audience of both parties involved creating mutually beneficial opportunities for exposure and revenue generation.

### 4. Secondary Market Impact
VADs may have implications for the secondary market of NFTs. Tokens associated with valuable VADs could see increased demand and potentially higher resale values. Collectors who missed out on early drops may be willing to pay a premium to acquire NFTs that offer special benefits. This secondary market activity can increase liquidity and trading volume within the NFT ecosystem, benefiting both creators and collectors.

### NFT documentation
NFT documentation is expected to make significant advances to meet the evolving needs of the NFT ecosystem. Better provenance and transparency will be prioritized, using technologies such as blockchain based provenance to establish an immutable record of NFTs ownership history. Cross chain compatibility will become more popular, enabling seamless migration and documentation across different blockchain networks. Secondly NFT documents will adopt more robust metadata standards, allowing for the inclusion of multimedia content, interactive features, and licensing information. The user interface of NFT markets and platforms will be improved to offer intuitive displays of documents, making it easier for collectors to access and understand the background and value of the token.

Legal and licensing considerations will also be included in the documentation, which will provide clarity about ownership rights and usage permissions. However, the integration of multimedia and interactive elements will create a more immersive documentary experience, revealing the unique features and characteristics of NFTs. As the NFT space continues to evolve, future NFT documentation will focus on transparency, functionality, and accessibility to provide a comprehensive and reliable system for obtaining and presenting the necessary NFT details.

### THE NFTs FUTURE OF GAMING
The video game industry sees great potential in the integration of NFTs as it could introduce a new audience and revolutionize the value attributed to digital objects. Currently gamers spend money on game keys, digital weapons, and rare cosmetic skins, although ownership of these items remains with game developers. However, NFTs have the potential to change this dynamic. the example of the growing interest in NFTs within gaming is seen in Counter Strike Global Offensive, where a gray market has formed around the trading of in game items. Some players have reportedly spent as much as $100,000 to get exclusive weapon skins. This illustrates the demand for unique and rare virtual items. Enter Blankos, a game that exploits the concepts of ownership and scarcity. Blankos allows players to collect, customize and sell NFTs of characters and items created by developers and well known brands. The initial success of Blankos is evident, with 100,000 NFTs purchased within a week of launching in June 2021. Notably, major brands and artists such as Burberry, Quix, and Deadmau5 have released exclusive items in the game, further showcasing the potential of NFTs.

This development indicates that NFTs have the potential to reshape the gaming industry by providing real ownership to players and enabling them to engage in digital commerce. Assets As more companies explore the integration of NFTs into their games, we can anticipate a significant shift in the way virtual items are perceived and valued.

### *CONCLUSION*
We have presented an overview of the multichain NFTs marketplace, highlighting the limitations of singlechain based architectures and the need for a decentralized platform that allows NFT transactions across multiple blockchain networks. The singlechain marketplace faced various challenges including performance issues, scalability limitations, and high transaction fees during high network usage. comprehensively explored the concept of NFT multichain architecture and discussed the challenges and opportunities in designing and implementing a multichain NFT marketplace. This architecture involves the integration of different blockchain networks that communicate with each other, enabling seamless NFT transactions. Explore the concept of mainchains interacting with sidechains, which establish a hierarchical structure connecting multiple blockchain networks. Moreover, Key challenges related to interoperability, security, scalability, and user adoption were identified in this context. To highlight these challenges, a new architecture for a multichain NFT marketplace was proposed. This architecture leverages the advantages of multiple blockchain networks and marketplaces, aiming to overcome identified bottlenecks. The proposed architecture was evaluated through a case study, demonstrating its ability to facilitate efficient and secure transactions across multiple blockchain networks.

For future work, we have planned to provide research in context of Multichain networks and explore vulnerability of architecture .and try to give best solution of these weaknesses.